\documentstyle[twocolumn,seceq]{jpsj}
\def\runtitle{Rodrigues Formula for the Nonsymmetric Multivariable
Hermite Polynomial}
\def\runauthor{Hideaki {\sc Ujino} and Miki {\sc Wadati}}
\def\Mae{\mbox{\hspace*{-20pt}}}

\def\Define{\mathop{\stackrel{\rm def}{=}}}
\newcommand{\myref}[1]{(\ref{#1})}
\newcommand{\vecvar}[1]{\mbox{\boldmath$#1$}}
\title{Rodrigues Formula for the Nonsymmetric Multivariable\\
Hermite Polynomial}
\author{Hideaki {\sc Ujino}%
\footnote{E-mail: ujino@monet.phys.s.u-tokyo.ac.jp} and Miki {\sc Wadati}}
\inst{Department of Physics, Graduate School of Science,
University of Tokyo,\\
Hongo 7--3--1, Bunkyo-ku, Tokyo 113--0033}
\recdate{\hspace{3cm}}
\abst{Applying a method developed by Takamura and Takano for the
nonsymmetric Jack polynomial, we present the Rodrigues formula for
the nonsymmetric multivariable Hermite polynomial.}
\kword{nonsymmetric multivariable Hermite polynomial,
Calogero model, Dunkl-Cherednik operator formulation, Rodrigues formula}
\begin{document}
\sloppy
\maketitle

\section{Introduction}\label{sec:introduction}
Explicit construction of the commuting conserved operators and
identification of their simultaneous eigenfunctions are important
fundamental problems in the study of quantum integrable systems.
The orthogonal basis spanned by the simultaneous eigenfunctions
of the conserved operators plays an essential role in detailed
studies on the integrable systems such as the calculation of the correlation
functions, the representation theory of the symmetry of the system and
so on. 

Among the integrable inverse-square interaction models in one dimension,
the Calogero and the Sutherland models~\cite{Calogero_1,Sutherland_1}
are considered to be the
``twins'' because the two models have their own Dunkl-Cherednik
operator formulations that share the same structure of the
commutator algebra.~\cite{Dunkl_1,Cherednik_1,Polychronakos_1,Ujino_1}
The Dunkl-Cherednik operator formulation provides an explicit construction
of the commuting conserved operators of the two models.
The celebrated Jack symmetric
polynomials~\cite{Jack_1,Stanley_1,Macdonald_1}
are the simultaneous eigenfunctions of conserved operators made of the
Cherednik operators of the Sutherland model. However, only a little had been
known about the symmetric simultaneous eigenfunctions of the conserved
operators of the Calogero model that are made of the Cherednik operators.%
~\cite{Ujino_2}
Motivated by the Rodrigues formula for the Jack symmetric polynomial that was
found by Lapointe and Vinet,~\cite{Lapointe_1,Lapointe_2}
we presented the Rodrigues formula for the Hi-Jack symmetric
(multivariable Hermite) polynomial and identified it as the simultaneous
eigenfunction of the conserved operators of the Calogero model.%
~\cite{Ujino_1,Ujino_3,Ujino_4} The multivariable Hermite polynomial is 
a one-parameter deformation of the Jack symmetric polynomial. They share
many common properties, which reflect the same algebraic structure of
the corresponding Dunkl-Cherednik operators.

To study the Calogero and Sutherland models including spin variables,
we need the nonsymmetric simultaneous eigenfunctions of the Cherednik
operators as the orthogonal basis of the orbital part of 
the wave function.~\cite{Kato_1,%
Takemura_1,Takemura_2}
Such a nonsymmetric
simultaneous eigenfunction of the conserved operators of the Sutherland
model is known to be the nonsymmetric Jack polynomial whose properties
are extensively studied in mathematical context. On the other hand,
the simultaneous eigenfunction of the Calogero model
is identified as the nonsymmetric multivariable Hermite
polynomial that is a one-parameter deformation of the nonsymmetric Jack
polynomial.
A recursive construction of the nonsymmetric Jack polynomial
was invented by Knop and Sahi.~\cite{Knop_1}
They also found out a combinatorial formula
of the nonsymmetric Jack polynomial which enables us to 
calculate the norm of the polynomial.~\cite{Sahi_1} Their results were
translated to the theory of the nonsymmetric multivariable
Hermite and Laguerre polynomials.~\cite{Baker_1} However, a simple expression
for an arbitrary nonsymmetric Jack polynomial and its variants that has a form
of successive operations of the raising operators to the vacuum
such as the Fock space of the quantum harmonic oscillator
has not been given in their formulation.

Quite recently, Takamura and Takano presented a simplified version of
an algebraic construction of the nonsymmetric Jack
polynomial.~\cite{Takamura_1}
Their intuitive formulation yields
a single expression for an arbitrary nonsymmetric Jack polynomial,
which is the advantage of Takamura and Takano's formulation to Knop
and Sahi's.~\cite{Knop_1}
Applying Takamura and Takano's results, we shall present the
Rodrigues formula for the nonsymmetric multivariable Hermite
polynomial. Recursion relations among norms and explicit forms of
norms for some simple cases will be also presented.

The outline of the paper is as follows. In \S\ref{sec:preparation},
we give a brief summary on the Dunkl-Cherednik operator formulation
for the Calogero model and the nonsymmetric multivariable Hermite 
polynomial. In \S\ref{sec:main}, the algebraic construction of
the nonsymmetric multivariable Hermite polynomial is presented in detail.
The final section is devoted to a summary.

\section{Dunkl-Cherednik Operators and Nonsymmetric Multivariable
Hermite Polynomials}\label{sec:preparation}
We give a brief summary on the Dunkl-Cherednik operator formulation
of the Calogero model and the nonsymmetric multivariable Hermite
polynomial. The Calogero Hamiltonian is expressed as
\begin{equation}
  \hat{H}_{\rm C} = \frac{1}{2}\sum_{j=1}^{N}
  \bigl(p_{j}^{2}+\omega^{2}x_{j}^{2}\bigr)
  +\frac{1}{2}\sum_{\stackrel{\scriptstyle j,k=1}{j\neq k}}^{N}
  \frac{a^{2}-aK_{jk}}{(x_{j}-x_{k})^{2}},
  \label{eqn:Calogero_Hamiltonian}
\end{equation}
where $p_{j} = -{\rm i}\frac{\partial}{\partial x_{j}}$ and
the coordinate exchange operator $K_{jk}$ is defined as
\[
  K_{jk}f(\cdots,x_{j},\cdots,x_{k},\cdots) = 
  f(\cdots,x_{k},\cdots,x_{j},\cdots).
\]
The ground state wave function is known to be
the real Laughlin wave function,
\[
  \phi_{\rm g}(\vecvar{x})
  = \prod_{1\leq j<k\leq N}|x_{j}-x_{k}|^{a}
  \exp\Bigl(-\frac{1}{2}\omega\sum_{j=1}^{N}x_{j}^{2}\Bigr),
\]
where the ground state energy is
\[
 E_{\rm g} = \frac{1}{2}N\omega\bigl(Na+(1-a)\bigr).
\]
The excited state wave function of the Calogero model is expressed as
the product of
an inhomogeneous polynomial and the ground state wave function.
To study the polynomial part of the wave functions, we introduce
a transformed Hamiltonian whose eigenfunctions are polynomials,
\begin{equation}
  H_{\rm C} \Define \bigl(\phi_{\rm g}(\vecvar{x})
  \bigr)^{-1}\bigl(\hat{H}_{\rm C}-E_{\rm g}\bigr)
  \phi_{\rm g}(\vecvar{x}).
  \label{eqn:transformed_Hamiltonian}
\end{equation}
In the following, we call eq.~\myref{eqn:transformed_Hamiltonian}
instead of eq.~\myref{eqn:Calogero_Hamiltonian} the Calogero Hamiltonian.
The commuting conserved operators for the Calogero model
are known to be the Cherednik operators. To show this,
we need to introduce the Dunkl operators and 
the creation-like and annihilation-like operators for
the Calogero model,
\begin{eqnarray*}
  & & \nabla_{j} \Define \frac{\partial}{\partial x_{j}}
  + a\sum_{\stackrel{\scriptstyle k=1}{k\neq j}}^{N}\frac{1}{x_{j}-x_{k}}
  (1-K_{jk}),\\
  & & \alpha_{l}^{\dagger}\Define
  x_{l}-\frac{1}{2\omega}\nabla_{l},\;\;\;\;
  \alpha_{l}\Define\nabla_{l}.
\end{eqnarray*}
Then the Cherednik operators are given by
\begin{equation}
  d_{l}\Define \alpha_{l}^{\dagger}\alpha_{l}
  +a\!\!\!\sum_{k=l+1}^{N}(K_{lk}-1)+a(N-l),\;\;\;\;
  [d_{l},d_{m}]=0.
  \label{eqn:Cherednik_op}
\end{equation}
In the above expression \myref{eqn:Cherednik_op},
we have adopted
Knop and Sahi's choice of the constant term in the Cherednik operator.
The commutator algebra among the Dunkl-Cherednik operators is listed as
follows,
\begin{eqnarray}
  \bigl[\alpha_{l}^{\dagger},\alpha_{m}^{\dagger}\bigr] & = & 0,\;\;
    \bigr[\alpha_{l},\alpha_{m}\bigr]=0,\nonumber\\
  \bigl[\alpha_{l},\alpha_{m}^{\dagger}\bigr] & = & \delta_{lm}(1+a
    \sum_{\stackrel{\scriptstyle k=1}{k\neq l}}^{N}K_{lk})
    -a(1-\delta_{lm})K_{lm},\nonumber\\
  \bigl[d_{l},\alpha_{m}^{\dagger}\bigr] & = & 
    \delta_{lm}\Bigl(\alpha_{l}^{\dagger}
    +a\sum_{k=1}^{l-1}\alpha_{l}^{\dagger}K_{lk}
    +a\!\!\!\sum_{k=l+1}^{N}\!\!\alpha_{k}^{\dagger}K_{lk}\Bigr)\nonumber\\
    & & -a\Bigl(\Theta(m-l)\alpha_{m}^{\dagger}K_{lm}
          + \Theta(l-m)\alpha_{l}^{\dagger}K_{lm}\Bigr),\nonumber\\
  \bigl[d_{l},\alpha_{m}\bigr] & = & 
    -\delta_{lm}\Bigl(-\alpha_{l}
    +a\sum_{k=1}^{l-1}\alpha_{k}K_{lk}
    +a\!\!\!\sum_{k=l+1}^{N}\!\!\alpha_{l}K_{lk}\Bigr)\nonumber\\
    & & +a\Bigl(\Theta(m-l)\alpha_{l}K_{lm}
          + \Theta(l-m)\alpha_{m}K_{lm}\Bigr),\nonumber\\
  \label{eqn:commutation_Dunkl_Cherednik}
\end{eqnarray}
where $\Theta(x)$ is the Heaviside function,
\[
  \Theta(x)=\left\{
    \begin{array}{ll}
      0, & x\leq 0, \\
      1, & x > 0.
    \end{array}\right.
\]
The Cherednik operators and the exchange operators satisfy
\begin{eqnarray}
  & & d_{l}K_{l,l+1}-K_{l,l+1}d_{l+1}=a,\;\;
      d_{l+1}K_{l,l+1}-K_{l,l+1}d_{l}=-a,\nonumber\\
  & & [d_{l},K_{m,m+1}]=0,\;\;(l\neq m,m+1).
  \label{eqn:commutation_Cherednik_exchange}
\end{eqnarray}
In terms of the Cherednik operators,
the Calogero Hamiltonian \myref{eqn:transformed_Hamiltonian}
can be expressed as
\[
  H_{\rm C}
  = \omega\sum_{l=1}^{N}\Bigl(d_{l}-\frac{1}{2}a(N-1)\Bigr).
\]
Thus the Cherednik operators $\{d_{l}|l=1,2,\cdots,N\}$
give a set of commuting conserved operators
of the Calogero model.

Hereafter, all the wave functions are labeled by 
the symbol $\lambda_{\sigma}$ that denotes a sequence of $N$ nonnegative
integers (composition) which is defined by a partition $\lambda$,
\begin{eqnarray*}
  & & \lambda\Define\{\lambda_{1}\geq\lambda_{2}\geq\cdots\geq
  \lambda_{N}\geq 0\},\\
  & & \lambda_{j}, \;\; j=1,2,\cdots,N,\mbox{ are nonnegative integers},
\end{eqnarray*}
and a distinct permutation $\sigma\in S_{N}$.
Distinct permutations $\sigma$ and $\tau$ for a partition
must satisfy
\[
  \lambda_{\sigma(j)}\neq\lambda_{\tau(j)}
\]
for some $j\in\{1,2,\cdots,N\}$.
For the definition of the nonsymmetric multivariable Hermite polynomial,
a partial order among compositions named the Bruhat order $\stackrel{\rm B}{<}$
is useful,
\[
  \mu_{\tau}\stackrel{\rm B}{<}\lambda_{\sigma}\Leftrightarrow
  \left\{
    \begin{array}{ll}
      1) & \mu\stackrel{\rm D}{<}\lambda, \\
      2) & \mbox{when $\mu=\lambda$ then
           the first}\\
         & \mbox{non-vanishing difference}\\
         & \tau(i)-\sigma(i)>0.
    \end{array}
  \right.
\]
Here, the symbol $\stackrel{\rm D}{<}$ denotes
the dominance order among partitions
\[
  \mu\stackrel{\rm D}{<}\lambda \Leftrightarrow \mu\neq\lambda,
  \; |\mu|=|\lambda|
  \mbox{ and }\sum_{k=1}^{l}\mu_{k}\leq\sum_{k=1}^{l}\lambda_{k},
\]
for all $l=1,2,\cdots,N$.
A set of indistinct permutations for a partition $\lambda$,
\[
  \{\sigma\}\Define\{\tau\in S_{N}|\lambda_{\tau}=\lambda_{\sigma}
  \mbox{ and }\lambda_{\tau}\stackrel{\rm B}{>}\lambda_{\sigma}
  \mbox{ for }\tau\neq\sigma\},
\]
is represented by the permutation $\sigma$ which gives the minimum in
the sense of the Bruhat order $\stackrel{\rm B}{<}$.

The nonsymmetric multivariable Hermite polynomial is the nondegenerate
simultaneous eigenfunction
of the Cherednik operators $\{d_{j}\}$ with the coefficient of its
top term $\vecvar{x}^{\lambda_{\sigma}}
\Define x_{1}^{\lambda_{\sigma(1)}}x_{2}^{\lambda_{\sigma(2)}}\cdots
x_{N}^{\lambda_{\sigma(N)}}$ conventionally taken to be unity,
\begin{eqnarray*}
  j_{\lambda_{\sigma}}(\vecvar{x};1/a,\omega) & = &
  \vecvar{x}^{\lambda_{\sigma}}
  +\!\!\!\!\!\!\!\!\!\!\sum_{\mu_{\tau}\stackrel{\rm B}{<}\lambda_{\sigma}\;
  {\rm or }\;|\mu|<|\lambda|}\!\!\!\!\!\!\!\!\!\!
  w_{\lambda_{\sigma}\mu_{\tau}}(a,\frac{1}{2\omega})
  \vecvar{x}^{\mu_{\tau}},\\
  d_{l}j_{\lambda_{\sigma}}(\vecvar{x};1/a,\omega) & = &
  \bar{\lambda}_{\sigma(l)}j_{\lambda_{\sigma}}(\vecvar{x};1/a,\omega),
\end{eqnarray*}
where
\begin{eqnarray*}
  \bar{\lambda}_{\sigma(l)} & \Define & \lambda_{\sigma(l)}
  -a\Bigl(\#\{1 \leq j < l |\lambda_{\sigma(j)}<\lambda_{\sigma(l)}\}\\
  & & +\#\{l < j \leq N |\lambda_{\sigma(j)}\leq\lambda_{\sigma(l)}\}\Bigr).
\end{eqnarray*}
The energy eigenvalue of the eigenfunction $j_{\lambda_{\sigma}}$
is given by
\[
  H_{\rm C}j_{\lambda_{\sigma}}
  = \sum_{k=1}^{N}\lambda_{\sigma(k)}j_{\lambda_{\sigma}}.
\]
In the following section, we shall construct
the simultaneous eigenfunction of the Cherednik operators.

\section{Rodrigues Formula}\label{sec:main}
Following the Takamura-Takano version of
the algebraic construction of the nonsymmetric Jack polynomials,
we shall present the algebraic
construction of the nonsymmetric multivariable Hermite polynomials.
Though the operator algebras of the Dunkl-Cherednik operators for the two
nonsymmetric polynomials share the same structure, we note that some
modifications, which will be explained later, are needed 
to apply Takamura and Takano's approach to the nonsymmetric
multivariable Hermite polynomials.

We introduce two types of operators to construct the nonsymmetric multivariable
Hermite polynomials. The first type is the quantum number exchange operator,
which is called the braid-exclusion operator in ref.~\citen{Takamura_1},
\begin{equation}
  X_{l,l+1}\Define {\rm i}[d_{l},K_{l,l+1}]=-{\rm i}[d_{l+1},K_{l,l+1}].
  \label{eqn:QN_exchange}
\end{equation}
We note that the quantum number exchange operator is Hermitian.
The quantum number exchange operators transpose the indices of the Cherednik
operators,
\begin{eqnarray}
  & & d_{l}X_{l,l+1}=X_{l,l+1}d_{l+1},\;\;
  d_{l+1}X_{l,l+1}=X_{l,l+1}d_{l},\nonumber\\
  & & [d_{k},X_{l,l+1}]=0,\;\;\;k\neq l,l+1.
  \label{eqn:commutation_Cherednik_QNE}
\end{eqnarray}
Thus they play the role of permutations in compositions.
By a straightforward calculation, we can verify the following relations
for the quantum number exchange operators,
\begin{eqnarray}
  & & \Mae X_{l,l+1}^{2} = (d_{l}-d_{l+1})^{2}-a^{2},\nonumber\\
  & & \Mae X_{l,l+1}X_{l+1,l+2}X_{l,l+1}=X_{l+1,l+2}X_{l,l+1}X_{l+1,l+2},
  \nonumber\\
  & & \Mae X_{l,l+1}X_{m,m+1}=X_{m,m+1}X_{l,l+1},\;\;(|l-m|\geq 2).
  \label{eqn:commutation_QNE}
\end{eqnarray}
The second type is
the raising operator that is given by
\begin{equation}
  a_{l}^{\dagger}\Define (X_{l,l+1}\cdots X_{N-1,N}e^{\dagger})^{l},\;\;\;\;
  l=1,\cdots,N,
  \label{eqn:raising_op}
\end{equation}
where $e^{\dagger}$ is a variant of the operator invented by Knop and Sahi,
\begin{equation}
  e^{\dagger}\Define K_{N,N-1}\cdots K_{2,1}\alpha_{1}^{\dagger}.
  \label{eqn:Knop-Sahi_op}
\end{equation}
What we want to show in the following is the commutation relation between
the Cherednik operator and the raising operator,
\begin{equation}
  [d_{l},a_{m}^{\dagger}]=\bigl(1-\Theta(l-m)\bigr)a_{m}^{\dagger}.
  \label{eqn:commutation_Cherednik_raising}
\end{equation}
We also show some other formulas to calculate the norms of the states
that will be algebraically constructed.

The Knop-Sahi operator $e^{\dagger}$ \myref{eqn:Knop-Sahi_op} satisfies
\begin{equation}
  e^{\dagger}e=d_{N},\;\; ee^{\dagger}=d_{1}+1,
  \label{eqn:motoha_unitarity}
\end{equation}
where $e$ is the Hermitian conjugate of the Knop-Sahi operator,
\[
  e=(e^{\dagger})^{\dagger}=\alpha_{1}K_{12}\cdots K_{N-1,N}.
\]
We note that the corresponding relations of the above expressions 
\myref{eqn:motoha_unitarity} for
the nonsymmetric Jack polynomials are simply
$e^{\dagger}e=ee^{\dagger}=1$.
Commutation relations related to the Knop-Sahi operator are
straightforwardly calculated,
\begin{eqnarray}
  & & d_{l}e^{\dagger}=e^{\dagger}d_{l+1},\;\; l=1,\cdots,N-1,\nonumber\\
  & & d_{N}e^{\dagger}=e^{\dagger}(d_{1}+1),\nonumber\\
  & & X_{l,l+1}e^{\dagger}=e^{\dagger}X_{l+1,l+2},\;\;
  l=1,\cdots,N-2,\nonumber\\
  & & X_{N-1,N}(e^{\dagger})^{2}=(e^{\dagger})^{2}X_{1,2}.
  \label{eqn:commutation_KSO}
\end{eqnarray}
We introduce
\begin{equation}
  b_{l}^{\dagger}\Define X_{l,l+1}\cdots X_{N-1,N}e^{\dagger}.
  \label{eqn:parts}
\end{equation}
Using eqs.~\myref{eqn:commutation_Cherednik_QNE},
\myref{eqn:commutation_QNE}, \myref{eqn:commutation_KSO}
and \myref{eqn:parts}, we easily verify
\begin{eqnarray}
  & & d_{l}b_{m}^{\dagger} = \left\{
  \begin{array}{ll}
    b_{m}^{\dagger}d_{l+1}, & 1\leq l \leq m-1,\\
    b_{m}^{\dagger}(d_{1}+1), & l=m,\\
    b_{m}^{\dagger}d_{l}, & m+1\leq l\leq N,
  \end{array}\right.\nonumber\\
  \lefteqn{X_{l,l+1}b_{m}^{\dagger}}\nonumber\\
  & & \;\;= \left\{
  \begin{array}{ll}
    b_{m}^{\dagger}X_{l,l+1}, & 1\leq l\leq m-1,\\
    b_{l+1}^{\dagger}\bigl((d_{l+1}-d_{1}-1)^{2}-a^{2}\bigr), & m=l,\\
    b_{l}, & m=l+1,\\
    b_{m}^{\dagger}X_{l+1,l+2}, & l+2\leq m\leq N,
  \end{array}\right.\nonumber\\
  & & b_{l}^{\dagger}b_{m}^{\dagger} = \left\{
  \begin{array}{ll}
    b_{m}^{\dagger}b_{l+1}^{\dagger}X_{1,2}, & l < m,\\
    X_{m-1,m}\cdots X_{l-2,l-1}(b_{l}^{\dagger})^{2}, & l\geq m.
  \end{array}\right.
  \label{eqn:commutation_parts}
\end{eqnarray}

Now we are ready to prove the commutation relation
\myref{eqn:commutation_Cherednik_raising}.
Using the definition of the raising operator \myref{eqn:raising_op},
we have
\[
  [d_{l},a_{m}^{\dagger}]
  =d_{l}(b_{m}^{\dagger})^{m}-(b_{m}^{\dagger})^{m}d_{l}.
\]
By use of eqs.~\myref{eqn:commutation_parts}, the first term of the above expression is cast into
\[
  d_{l}(b_{m}^{\dagger})^{m}=\left\{
  \begin{array}{ll}
    (b_{m}^{\dagger})^{m}d_{l}, & l>m,\\
    (b_{m}^{\dagger})^{m}(d_{l}+1), & l\leq m.
  \end{array}\right.
\]
Thus we obtain
\[
  \bigl[d_{l},a_{m}^{\dagger}\bigr] =\left\{
  \begin{array}{ll}
    0 , & l>m,\\
    a_{m}^{\dagger}, & l\leq m,
  \end{array}\right.
\]
which proves eq.~\myref{eqn:commutation_Cherednik_raising}.
We should note that the raising operators are commutative
with each other,
\[
  [a_{l}^{\dagger},a_{m}^{\dagger}]=0,
\]
as is shown from eqs.~\myref{eqn:commutation_parts}.

The following relations are useful when we calculate the norms 
of the eigenstates,
\begin{eqnarray}
  & & b_{l}^{\dagger}b_{l} = d_{l}\prod_{k=l+1}^{N}
  \bigl((d_{l}-d_{k})^{2}-a^{2}\bigr),\nonumber\\
  & & b_{l}b_{l}^{\dagger} = (d_{1}+1)\prod_{k=l+1}^{N}
  \bigl((d_{1}-d_{k}+1)^{2}-a^{2}\bigr),\nonumber\\
  & & b_{N}^{\dagger}b_{N} = d_{N},\;\;b_{N}b_{N}^{\dagger}=d_{1}+1.
  \label{eqn:number-like_parts}
\end{eqnarray}
Using eqs.~\myref{eqn:commutation_Cherednik_exchange},
\myref{eqn:commutation_parts} and \myref{eqn:number-like_parts},
we can rewrite the number-like operators in terms
of the Cherednik operators,
\begin{eqnarray}
  a_{l}^{\dagger}a_{l} & = & \prod_{m=1}^{l}\Bigl(
  \prod_{k=l+1}^{N}(d_{m}-d_{k})^{2}-a^{2}\Bigr)d_{m},\nonumber\\
  a_{l}a_{l}^{\dagger} & = & \prod_{m=1}^{l}\Bigl(
  \prod_{k=l+1}^{N}(d_{m}-d_{k}+1)^{2}-a^{2}\Bigr)(d_{m}+1),
  \nonumber\\
  \label{eqn:number-like}
\end{eqnarray}
for $l=1,\cdots,N$.
These relations are different from the corresponding formulas
for the nonsymmetric Jack polynomials, which causes a slight modification
of the norms.
We also need the following formulas in the calculation of the norms,
\begin{eqnarray}
  & & X_{l,l+1}a_{m}^{\dagger} = a_{m}^{\dagger}X_{l,l+1},\;\;
  m\neq l,\nonumber\\
  & & a_{l}^{\dagger}X_{l,l+1}a_{l}^{\dagger}=
  \bigl((d_{l}-d_{l+1}-1)^{2}-a^{2}\bigr)X_{l,l+1}b_{l-1}^{\dagger}
  b_{l+1}^{\dagger},\nonumber\\
  \label{eqn:commutation_number-like_QNE}
\end{eqnarray}
which can be verified by use of eqs.~\myref{eqn:commutation_parts}.

We are ready to write down the Rodrigues formula for the nonsymmetric
multivariable Hermite polynomials upto normalization.
For the composition $\lambda_{\sigma}$ where the distinct permutation
$\sigma$ is expressed by the product of transpositions as
\[
  \sigma=(k_{l},k_{l}+1)\cdots(k_{2},k_{2}+1)(k_{1},k_{1}+1),
\]
we can show by use of eqs.~\myref{eqn:commutation_Cherednik_QNE} and 
\myref{eqn:commutation_Cherednik_raising} that
the nonsymmetric polynomial $k_{\lambda_{\sigma}}$ generated by
the Rodrigues formula,
\begin{eqnarray}
  k_{\lambda_{\sigma}} & \Define & X_{k_{1},k_{1}+1}X_{k_{2},k_{2}+1}\cdots
  X_{k_{l},k_{l}+1}\nonumber\\
  & & (a_{1}^{\dagger})^{\lambda_{1}-\lambda_{2}}
  (a_{2}^{\dagger})^{\lambda_{2}-\lambda_{3}}\cdots
  (a_{N}^{\dagger})^{\lambda_{N}}\cdot 1,
  \label{eqn:Rodrigues}
\end{eqnarray}
satisfies the definition of the nonsymmetric multivariable Hermite
polynomials $j_{\lambda_{\sigma}}$ except for the normalization,
\[
  d_{j}k_{\lambda_{\sigma}}=\bar{\lambda}_{\sigma(j)}k_{\lambda_{\sigma}}.
\]
Thus we conclude $k_{\lambda_{\sigma}}\propto j_{\lambda_{\sigma}}$.
Since the Cherednik operators ${d_{l}}$ are Hermitian with respect to
the following conventional inner product,
\begin{eqnarray*}
  \langle f,g\rangle & \Define & \int_{-\infty}^{\infty}\prod_{j=1}^{N}
  {\rm d}x_{j}|\phi_{\rm g}|^{2}f^{\dagger}(\vecvar{x})g(\vecvar{x}),\\
  |f|^{2} & \Define & \langle f,f\rangle,
\end{eqnarray*}
the nonsymmetric multivariable Hermite polynomials are orthogonal
with respect to the inner product,
\[
  \langle k_{\lambda_{\sigma}},k_{\mu_{\tau}}\rangle = 
  |k_{\lambda_{\sigma}}|^{2}
  \delta_{\lambda_{\sigma},\mu_{\tau}}\Leftrightarrow
  \langle j_{\lambda_{\sigma}},j_{\mu_{\tau}}\rangle=
  |j_{\lambda_{\sigma}}|^{2}
  \delta_{\lambda_{\sigma},\mu_{\tau}}.
\]
For the case $\sigma={\rm id}$, the norm
$\langle k_{\lambda},k_{\lambda}\rangle\Define|k_{\lambda}|^{2}$
is calculated in an algebraic
manner,
\begin{eqnarray*}
  \frac{\langle k_{\lambda},k_{\lambda}\rangle}{\langle 1,1\rangle}
  & = & \prod_{k=1}^{N}\prod_{l=1}^{k}
  \prod_{m=1}^{\lambda_{k}-\lambda_{k+1}}
  \Bigl(\lambda_{k}+a(N-l)-(m-1)\Bigr)\\
  & & \!\!\!\prod_{n=k+1}^{N}\!\!\!
  \Bigl((\lambda_{k}-\lambda_{n}+a(n-l)-(m-1))^{2}
  -a^{2}\Bigr),
\end{eqnarray*}
where $\langle 1,1\rangle$ is the vacuum normalization,~\cite{Diejen_1}
\[
  \langle 1,1\rangle
  = \frac{(2\pi)^{\frac{N}{2}}}{(2\omega)^{\frac{1}{2}N(Na+(1-a))}}
  \prod_{1\le j\le N}\frac{\Gamma(1+ja)}{\Gamma(1+a)},
\]
with $\Gamma(z)$ being the gamma function.
We note that the above formula for the ratio of the norms of
$k_{\lambda_{\sigma}}$ and the ground state is different from the
corresponding formula in Takamura and Takano's result. The difference
comes from the definitions of the inner products that
respectively make the nonsymmetric Jack and nonsymmetric multivariable
Hermite polynomials orthogonal.
For the general case, the norm
$\langle k_{\lambda_{\sigma}},k_{\lambda_{\sigma}}\rangle
\Define|k_{\lambda_{\sigma}}|^{2}$ satisfies
the following recursion relation,
\[
  \frac{\langle X_{l,l+1}k_{\lambda_{\sigma}},%
  X_{l,l+1}k_{\lambda_{\sigma}}\rangle}%
  {\langle k_{\lambda_{\sigma}},k_{\lambda_{\sigma}}\rangle}
  =\Bigl((\bar{\lambda}_{\sigma(l)}-\bar{\lambda}_{\sigma(l+1)})^{2}
  -a^{2}\Bigr),
\]
which is equivalent to
\[
  X_{l,l+1}\frac{k_{\lambda_{\sigma}}}{|k_{\lambda_{\sigma}}|}=
  \Bigl((\bar{\lambda}_{\sigma(l)}-\bar{\lambda}_{\sigma(l+1)})^{2}
  -a^{2}\Bigr)^{\frac{1}{2}}
  \frac{k_{\lambda_{\sigma(l,l+1)}}}{|k_{\lambda_{\sigma(l,l+1)}}|}.
\]
The coefficient of the r.h.s.~of the above expression becomes zero
when the two permutations $\sigma$ and $\sigma(l,l+1)$ are indistinct
with respect to the Young diagram $\lambda$,
$\lambda_{\sigma(l)}=\lambda_{\sigma(l+1)}$.
Explicit form for $\sigma={\rm id}$ and recursion relations of the norms
can be straightforwardly derived from eqs.~\myref{eqn:commutation_QNE},
\myref{eqn:commutation_parts} and \myref{eqn:number-like}.

\section{Summary}\label{sec:summary}
Generalizing the method developed by Takamura and Takano for the nonsymmetric
Jack polynomials, we have presented an algebraic construction of the
nonsymmetric multivariable Hermite polynomials which span the orthogonal
basis of the orbital part of the wave function of the Calogero model including
spin variables. Because of the difference in the definitions
of the inner product, the formulas of the norms of the two nonsymmetric
orthogonal polynomials are different from each other.

Though the Calogero and the Sutherland models share a lot of
properties of their algebraic structures and structures of their
Hilbert spaces, calculations of correlation functions have been done
only for the Sutherland model.~\cite{Ha_1,Ha_2,Uglov_1}
The difficulty in the calculation toward the correlation functions
of the Calogero model comes from the difference of the analytic
properties and the inner product of the nonsymmetric multivariable
Hermite polynomial and those of the nonsymmetric Jack polynomial.
Thus any new approach and formulation for the Calogero and the Sutherland
models should be tested if they will give a breakthrough in the
calculation of correlation functions of the two models on an equal basis.
We need further efforts toward such directions.

Takamura and Takano's approach
is intuitive enough to give a simple expression
for arbitrary nonsymmetric multivariable Hermite polynomial.
The coefficient of the top term $\vecvar{x}^{\lambda_{\sigma}}$ of
the polynomial $k_{\lambda_{\sigma}}$, however,
has not been computed for general cases, which is a disadvantage
to Knop and Sahi's formulation.
The same approach will be also applicable to
the nonsymmetric multivariable Laguerre polynomial with even parity,
which span the orthogonal basis of the $B_{N}$ Calogero
model~\cite{Yamamoto_1}. These problems are also left for future studies.

\section*{Acknowledgements}
The authors are grateful to Dr.~A.~Takamura for sending us the 
paper~\cite{Takamura_1} prior to publication.

One of the authors (HU) appreciates the Research Fellowships of the
Japan Society for the Promotion of Science for Young Scientists.

\end{document}